\documentclass[epj]{svjour}
\usepackage{graphics}
\begin{document}
\title{Self-learning Mutual Selection Model for Weighted Networks}
\author{Jian-Guo Liu\inst{1},\thanks{\emph{Present address:} liujg004@yahoo.com.cn}
Yan-Zhong Dang\inst{1}, Wen-Xu Wang\inst{2}, Zhong-Tuo Wang
\inst{1}, Tao Zhou\inst{2}, Bing-Hong Wang\inst{2}, Qiang Guo
\inst{3}, Zhao-Guo Xuan \inst{1},
Shao-Hua Jiang \inst{1} and Ming-Wei Zhao\inst{1} 
%
}                     
\offprints{}          
\institute{Institute of System Engineering, Dalian University of
Technology, Dalian Liaoning, 116023, P R China \and Department of
Modern Physics, University of Science and Technology of China, Hefei
Anhui, 230026, P R China \and School of Science,
                         Dalian Nationalities University,
                         Dalian Liaoning, 116600, P R China}
\date{Received: date / Revised version: date}
%
\abstract{In this paper, we propose a self-learning mutual selection
model to characterize weighted evolving networks. By introducing the
self-learning probability $p$ and the general mutual selection
mechanism, which is controlled by the parameter $m$, the model can
reproduce scale-free distributions of degree, weight and strength,
as found in many real systems. The simulation results are consistent
with the theoretical predictions approximately. Interestingly, we
obtain the nontrivial clustering coefficient $C$ and tunable degree
assortativity $r$, depending on the parameters $m$ and $p$. The
model can unify the characterization of both assortative and
disassortative weighted networks. Also, we find that self-learning
may contribute to the assortative mixing of social networks.
} 
\maketitle
\section{Introduction}
\label{intro} In recent years, many empirical findings have
triggered the devotion of research communities to understand and
characterize the evolution mechanisms of complex networks including
the Internet, the World-Wide Web, the scientific collaboration
networks and so on\cite{WS98,BA99,AB02,DM02,New,XFWang01}. Many
empirical evidences indicate that the networks in various fields
have some common characteristics. They have a small average distance
like random graphs, a large clustering coefficient and power-law
degree distribution \cite{WS98,BA99}, which is called the
small-world and scale-free characteristics. Recent works on the
complex networks have been driven by the empirical properties of
real-world networks and the studies on network dynamics
\cite{ZCP,ZT,ZM,YHJ}. The first successful attempt to generate
networks with large clustering coefficient and small average
distance is that of Watts and Strogatz (WS model) \cite{WS98}.
Another significant model proposed by Barab\'{a}si and Albert is
called scale-free network (BA network) \cite{BA99}. The BA model
suggests that growth and preferential attachment are two main
self-organization mechanisms of the scale-free networks structure.
However, the real systems are far from boolean structure. The purely
topological characterization will miss important attributes often
encountered in real systems. Most recently, the access to more
complete empirical data allows scientists to study the weight
evolution of many real systems. This calls for the use of weighted
network representation. The weighted network is often denoted by a
weighted adjacency matrix with element $w_{ij}$ representing the
weight on the edge connecting node $i$ and $j$. As a note, this
paper will only consider undirected graphs where weights are
symmetric, i.e. $w_{ij}=w_{ji}$. The strength $s_i$ of node $i$ is
usually defined as $s_i=\sum_{j\in \Gamma(i)}w_{ij}$, where the sum
runs over the neighbor node set $\Gamma(i)$. But in some cases, the
sum can not reflect the node strength completely. Take the
scientific collaboration networks for example, the strength of a
scientist include the publications collaborated with others and the
publications written only by himself or herself. Inspired by this
idea, the node strength is defined as $s_i=\sum_{j\in
\Gamma(i)}w_{ij}+\eta_i$, where $\eta_i$ is node $i$'s
self-attractiveness. As confirmed by the empirical data, complex
networks exhibit power-law distributions of degree $P(k)\sim
k^{-\gamma}$ with $2\leq\gamma \leq 3$ \cite{EPJB,15} and weight
$P(w)\sim w^{-\theta}$ \cite{WX}, as well as strength $P(s)\sim
s^{-\alpha}$ \cite{15}. The strength usually reveals scale-free
property with the degree $s\sim k^{\beta}$, where $\beta>1$
\cite{15,13,14}. Driven by new empirical findings, Barrat {\it et
al.} have presented a simple model (BBV for short) to study the
dynamical evolution of weighted networks \cite{152}. But its
disassortative property can not answer the open question: why social
networks are different from other disassortative ones? Previous
models can generate either assortative networks \cite{17,18,19} or
disassortative ones \cite{15,17,18,WWX2,WWX}. 
Our work may shed some new light to answer the question: is there a
generic explanation for the difference of assortative and
disassortative networks.

Previous network models often adopt the mechanism that only newly
added nodes could select the pre-existing nodes according to the
preferential mechanism. However, the evolution picture ignores the
fact that old nodes will choose the young nodes at the same time.
Inspired by this idea, Wang {\it et al.} have presented the mutual
selection mechanism, which leads to the creation and reinforcement
of connections \cite{WWX2}. But the model ignored the fact that
every node would enhance its strength not only by creating new links
to others, but also could by self-learning. In this paper,
self-learning means that a node enhances its strength only by itself
without creating new links to others. Inspired by this idea, we
propose a weighted network model that considers the topological
evolution under the general mechanisms of mutual selection and
self-learning. It can mimic the evolution of many real-world
networks. Our microscopic mechanisms can well explain the
characteristics of scale-free weighted networks, such as the
distributions of degree, weight and strength, as well as the
nonlinear strength-degree correlation, nontrivial clustering
coefficient, assortativity coefficient and hierarchical structure
that have been empirically observed \cite{EPJB,15,6,7,Li,PP}. Also,
the model appears as a more general one that unifies the
characterization of both assortative and disassortative weighted
networks.

\section{Construction of the model}
\label{sec:1}

Our model is defined as following. The model starts from $N_0=m$
isolated nodes, each with an initial attractiveness $s_0$. In this
paper, $s_0$ is set to be 1.
\begin{description}
\item[(i)] At each time step, a new node with strength $s_0$ and
degree zero is added in the network;

\item[(ii)] Every node strength of the
network would increase by 1 with the probability $p$; According to
the probability $(1-p)$, each existing node $i$ selects $m$ other
existing nodes for potential interaction according to the
probability Equ. (\ref{F1.1}). Here, the parameter $m$ is the number
of candidate nodes for creating or strengthening connections, $p$ is
the probability that a node would enhance $\eta_i$ by 1.
\end{description}
\begin{equation}\label{F1.1}
\Pi_{i\rightarrow j}=\frac{s_j}{\sum_k s_k-s_j}.
\end{equation}
where $s_i=\sum_{j\in \Gamma(i)}w_{ij}+\eta_i$. If a pair of
unlinked nodes is mutually selected, then an new connection will be
built between them. If two connected nodes select each other, then
their existing connection will be strengthened, i.e., their edge
weight will be increased by 1. We will see that $m$ and $p$ control
the evolution of our network.

The evolution of real-world network can be easily explained by our
model mechanisms. Take the scientific collaboration networks as an
example: the collaboration of scientists requires their mutual
status and acknowledgements. A scientists would like to collaborate
with others, whom have strong scientific potentials and long
collaborating history. On the contrary, he may write paper or
publications only by himself. When he publishes paper as the sole
author, his strength also increases, which can be reflected by
$\eta$. For technological networks with traffic taking place on
them, both the limit of resources and the internal demand of traffic
increment for keeping the normal function of the networks may cause
the mutual selections.

\begin{figure}
\resizebox{0.54\textwidth}{!}{%
  \includegraphics{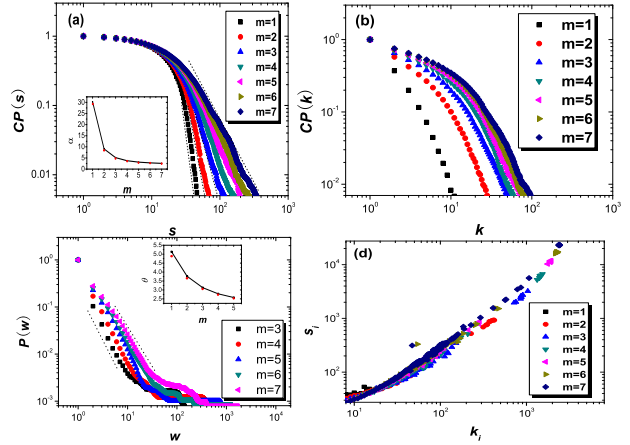}
} \caption{(Color online) Numerical results by choosing $p=0.004$.
The data are averaged over ten
       independent runs of network size $N=7000$. (a)Cumulative probability strength distribution $CP(s)$ with various
       $m$. The results are consistent with a power-law distribution $CP(s)\sim s^{\alpha}$. The inset reports the
       obtained values by data fitting (full circles) in comparison with the theoretical prediction $\alpha=2+
       z/[(1-p)^2m^2]$ (line). (b) Cumulative probability degree distribution $CP(k)$ with various $m$, which
       demonstrates that the degree distributions have power-law tail. (c) Cumulative probability weight distributions with
       various $m$, which are consistent with the power-law tail $P(w)\sim w^{\theta}$. (d) To different $m$, the average
       strength $s_i$ of nodes with connectivity $k_i$. We observe the nontrivial strength-degree correlation
       $s\sim k^{\beta}$ in the log-log scale.}
\label{fig:1}       
\end{figure}

\begin{figure}
\resizebox{0.54\textwidth}{!}{%
  \includegraphics{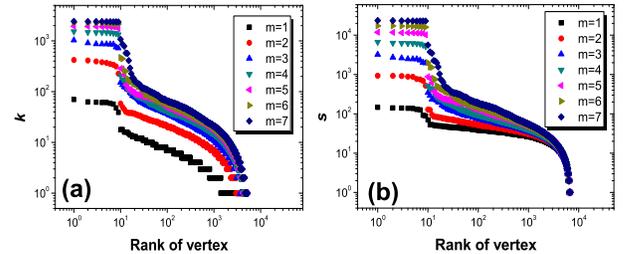}
} \caption{(Color online) Zipf plot of the degree and node strength
when $p=0.004$.}
\label{fig:2}       
\end{figure}

\begin{figure}
\resizebox{0.54\textwidth}{!}{%
  \includegraphics{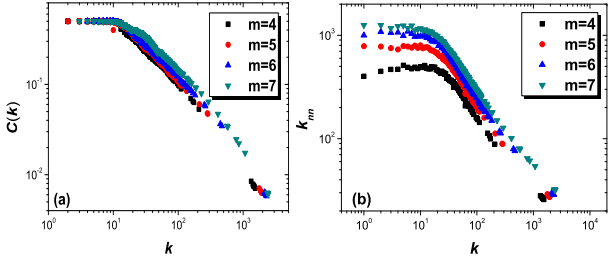}
} \caption{(Color online) The scale of $C(k)$ and $k_{nn}$ with $k$
for various $m$ when $p=0.004$. The data are averaged over 10
independent runs of network size $N=7000$.}
\label{fig:3}       
\end{figure}

\begin{figure}
\resizebox{0.54\textwidth}{!}{%
  \includegraphics{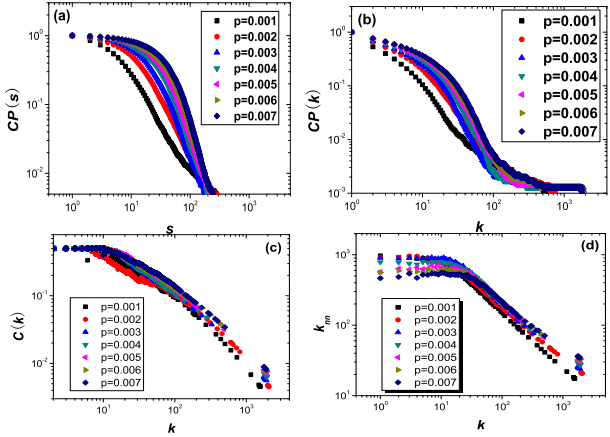}
} \caption{(Color online) Numerical results by choosing $m=5$ with
various $p$. The data are averaged over ten
       independent runs of network size $N=7000$. (a)Cumulative probability strength distributions $CP(s)$ with various
       $p$. The results are consistent with a power-law distribution $CP(s)\sim s^{\alpha}$. (b) Cumulative probability
       degree distributions $CP(k)$ with various $p$, which
       demonstrate that the degree distributions have power-law tail. (c) The clustering coefficient $C(k)$ depending
on connectivity $k$ for various $p$. (d) Average connectivity
$k_{nn}$ of the nearest neighbors of a node depending on its
connectivity $k$ for different $p$.}
\label{fig:4}       
\end{figure}

\begin{figure}
\resizebox{0.54\textwidth}{!}{%
  \includegraphics{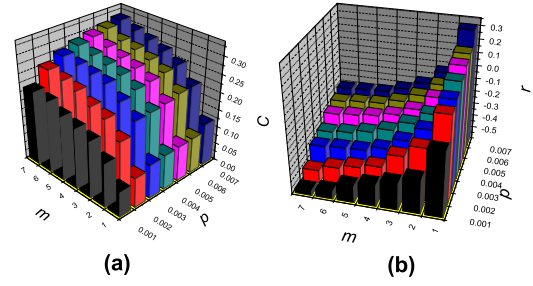}
} \caption{(Color online) The scale of $C$ and $r$ with various $m$
and $p$. The data are averaged over 10 independent runs of network
size $N=7000$.}
\label{fig:5}       
\end{figure}

\section{characteristics of the model}

Considering the rule that $w_{ij}$ is updated only if node $i$ and
$j$ select each other, and using the continuous approximation, then
the time evolution of weight can be computed analytically as
$$
 \frac{dw_{ij}}{dt} =  \frac{(1-p)ms_j}{\sum_{k(\neq
i)}s_k}\cdot  \frac{(1-p)ms_i}{\sum_{k(\neq i)}s_k}
$$
\begin{equation}\label{F2.2}
        \approx  \frac{(1-p)^2m^2(s_is_j)}{(\sum_{k}s_k)^2}.
\end{equation}
Hence, the strength $s_i(t)$ is updated by this rate
\begin{equation}\label{F2.3}
\frac{ds_i}{dt}=\sum_j\frac{dw_{ij}}{dt}+p\approx
\frac{(1-p)^2m^2s_i}{\sum_{k}s_k}+p.
\end{equation}
Notice that
$$
\sum_i s_i=\int^t_0\frac{\sum_i
ds_i}{dt}dt=\int^t_0\sum_i\Big[\frac{(1-p)^2m^2s_i}{\sum_k
s_k}+p\Big]dt.
$$
Thus, Equ. (\ref{F2.3}) can be expressed by
\begin{equation}\label{F2.5}
\frac{ds_i}{dt}=\frac{(1-p)^2m^2s_i}{(1-p)^2m^2t+pt^2}+p.
\end{equation}
When $p\sim O(N^{-1})$, the solution of Equ. (\ref{F2.5}) can be
obtained approximately as follows
\begin{equation}\label{F2.52}
s_i(t)\sim t^{\lambda},
\end{equation}
where
$$
\lambda=\frac{(1-p)^2m^2}{(1-p)^2m^2+z},
$$
 and $z=pN$ is a constant.
Then, we can get that the strength distribution obeys the power-law
$P(s)\sim s^{-\alpha}$ \cite{152} with exponent
\begin{equation}\label{2.6}
\alpha=1+\frac{1}{\lambda}=2+\frac{z}{(1-p)^2m^2}.
\end{equation}
One can also obtain the evolution behavior of the weight
distribution $P(w)\sim w^{\theta}$ \cite{22}, where
\begin{equation}
\theta= 2 + \frac{2z}{(1-p)^2m^2-z},
\end{equation}
and the degree distribution $P(k)\sim k^{-\gamma}$, where
\begin{equation}
\gamma= 2+\frac{z}{(1-p)^2m^2}.
\end{equation}
By choosing different values of $p$ and $m$, we perform numerical
simulations of networks which is consistent with the theoretical
predictions. Fig. 1(a)-(d) present the probability distributions of
strength, degree and weight, as well as the strength-degree
correlation, fixed $p = 0.004$ and tuned by $m$. Fig. 1(a) gives the
node strength distribution $P(s)\sim  s^{\alpha}$, which is in good
agreement with the theoretical expression Equ. (\ref{2.6}). Fig.
1(b) gives the node degree distribution $P(k)\sim k^{-\gamma}$. Fig.
1(c) reports the probability weight distribution, which also shows
the power-law behavior $P(w)\sim w^{\theta}$. Fig. 1(d) reports the
average strength of nodes with degree $k_i$, which displays a
nontrivial power-law behavior $s\sim k^{\beta}$ as confirmed by
empirical measurements \cite{15}. Fig. 2(a)-(b) show the Zipf plot
of the simulation results by fixing a moderate value $p=0.004$ and
varying $m$.  Fig. 2(a) confirms with the math collaboration network
and the Zipf plot of Fig. 1(a) in Ref. \cite{Li}. Fig. 3(a)-(b) give
the clustering coefficient $C(k)$ depending on connectivity $k$ and
the average connectivity $k_{nn}$ of the nearest neighbors of a node
for various $m$. From the numerical results, we can obtain the
conclusion that the larger the probability $p$, the larger the
effect of exponential correction at the head. However, the power-law
tail which again recovers the theoretical exponent expressions can
still be observed. Fig. 4 gives the numerical results for various
$p$ when $m=5$.

Depending on the parameters $p$ and $m$, the unweighted clustering
coefficient $C$, which describes the statistic density of connected
triples, and degree assortativity $r$ \cite{r,r2} are demonstrated
in Fig. 5. The assortative coefficient $r$ can be calculated from
\begin{equation}
r=\frac{M^{-1}\sum_i j_i
k_i-[M^{-1}\sum_i\frac{1}{2}(j_i+k_i)]^2}{M^{-1}\sum_i\frac{1}{2}(j_i^2+k_i^2)-[M^{-1}\sum_i\frac{1}{2}(j_i+k_i)]^2},
\end{equation}
where $j_i$, $k_i$ are the degrees of the vertices at the ends of
the $i$th edge, with $i=1, 2, \cdots,M$. From Fig. 5(a), we can find
that $C$ for fixed $m$ increases with $p$ slightly, and $C$ for
fixed $p$ monotonously increases with $m$. The clustering
coefficient of our model is tunable in a broad range by varying both
$m$ and $p$, which makes it more powerful in modelling real-world
networks. As presented in Fig. 5(b), degree assortativity $r$ for
fixed $p$ decreased with $m$, unlike the clustering case. While $r$
for fixed $m$ increases with $p$ slightly. The model can generates
disassortative networks for small $m$ and large $p$, which can best
mimic technological networks. At large $p$ and small $m$,
assortative networks emerge and can be used to mimic social
networks, such as the scientific collaboration networks. In the
model, enhancing the probability $p$ can be considered as the
probability that a node would like to study by itself to enhance its
attractiveness or prestige. In the competitive social networks, all
nodes face many competitors. In order to subsist or gain
honorableness, they must enhance their attractiveness or ability by
studying themselves or collaborating with others. This explains the
origin of assortative mixing in our model and may shed light on the
open question: why social networks are different from other
disassortative ones? For example, in the scientific collaboration
networks, the attractiveness of a scientist could not be represented
simple by the publications collaborated with others. Indeed, there
are many other important qualities that will contribute to the
attractiveness of a scientist, for instance, the publications
written by himself, etc. Perhaps the different self-learning
probability contributes to human beings fundamental differences. On
the other hand, $m$ indicates the interaction frequency among the
network internal components. If $m$ increases, the hubs would link
more and more ``young" sites. Thus, the reason why the
disassortativity of the model is more sensitive to $m$ lies in that
collaboration is more important than self-learning in the
technological networks. In addition, the components of technological
networks are usually physical devices, which can not study by
itself. Combining these two parameters together, two competitive
ingredients, which may be responsible for the mixing difference in
real complex networks, are integrated in our model.

\section{Conclusion and Discussion}
In summary, integrating the mutual selection mechanism between nodes
and the self-learning mechanism, our network model provides a wide
variety of scale-free behaviors, tunable clustering coefficient, and
nontrivial degree-degree and strength-degree correlations, just
depending on the probability of self-learning $p$ and the parameter
$m$ which governs the total weight growth. All the statistic
properties of our model are found to be supported by various
empirical data. Interestingly and specially, studying the
degree-dependent average clustering coefficient $C(k)$ and the
degree-dependent average nearestneighbors¡¯ degree $k_{nn}(k)$ also
provides us with a better description of the hierarchies and
organizational architecture of weighted networks. Our model may be
beneficial for future understanding or characterizing real networks.

Due to the apparent simplicity of our model and the variety of
tunable results, we believe our present model, for all practical
purposes, might be applied in future weighted network research.

This work has been supported by the Chinese Natural Science
Foundation of China under Grant Nos. 70431001, 70271046 and
70471033.

\end{document}